\documentclass[conference]{IEEEtran}
\IEEEoverridecommandlockouts
\usepackage{cite} 
\usepackage{amsmath,amssymb,amsfonts,amsthm}
\usepackage{algorithmic}
\usepackage{graphicx}
\usepackage{textcomp}
\usepackage{xcolor}
\def\BibTeX{{\rm B\kern-.05em{\sc i\kern-.025em b}\kern-.08em
    T\kern-.1667em\lower.7ex\hbox{E}\kern-.125emX}}
\usepackage{comment}
\usepackage[utf8]{inputenc}
\usepackage[T1]{fontenc}
\usepackage{hyperref}

\usepackage{stmaryrd}
\newcommand{\semantics}[1]{\llbracket #1 \rrbracket}
\newcommand{\types}{\mathcal T}
\newcommand{\constants}{\mathcal C}
\newcommand{\predicates}{\mathcal P}
\newcommand{\variables}{\mathcal V}
\newcommand{\functions}{\mathcal F}
\newcommand{\interpretation}{\mathcal I}
\newcommand{\objects}{\mathcal O}

\theoremstyle{remark}

\usepackage[inline,nomargin,index,marginclue]{fixme}
\fxusetheme{color}
\FXRegisterAuthor{ap}{aap}{AP}

\newcommand{\ltlU}{\ \mathsf{U}\ }
\newcommand{\ltlX}{\ \mathsf{X}\ }
\newcommand{\ltlS}{\ \mathsf{S}\ }
\newcommand{\ltlXm}{\ \mathsf{X}^{-1}\ }
\newcommand{\ltlF}{\ \mathsf{F}\ }
\newcommand{\ltlFm}{\ \mathsf{F}^{-1}\ }
\newcommand{\ltlG}{\mathsf{G}}

\newcommand{\ltlR}{\ \mathsf{R} \ }

\newcommand{\multiletterinmath}[1]{\mathit{#1}}

\newcommand{\AT}{\multiletterinmath{AT}}
\newcommand{\RG}{\multiletterinmath{RG}}
\newcommand{\TG}{\multiletterinmath{TG}}
\newcommand{\TF}{\multiletterinmath{TF}}
\newcommand{\ENV}{\multiletterinmath{ENV}}
\newcommand{\SE}{\multiletterinmath{SE}}
\newcommand{\TTES}{\multiletterinmath{TTES}}
\newcommand{\TES}{\multiletterinmath{TES}}

\newcommand{\vto}{\multiletterinmath{to}}
\newcommand{\vrt}{\multiletterinmath{rt}}
\newcommand{\vtrj}{\multiletterinmath{trj}}

\newcommand{\vrm}{\multiletterinmath{rm}}
\newcommand{\vth}{\multiletterinmath{th}}
\newcommand{\vst}{\multiletterinmath{st}}
\newcommand{\vbr}{\multiletterinmath{br}}

\newcommand{\vloc}{\multiletterinmath{loc}}
\newcommand{\vlbl}{\multiletterinmath{lbl}}
\newcommand{\vlen}{\multiletterinmath{len}}
\newcommand{\vdst}{\multiletterinmath{dst}}
\newcommand{\vsrc}{\multiletterinmath{src}}
\newcommand{\vrts}{\multiletterinmath{rts}}
\newcommand{\vtrjs}{\multiletterinmath{trjs}}
\newcommand{\vpr}{\multiletterinmath{pr}}
\newcommand{\vprs}{\multiletterinmath{prs}}

\begin{document}

\title{Leveraging Compositional Methods for Modeling and Verification of an Autonomous Taxi System\\
\thanks{This material is based upon work supported by the National Science Foundation under NSF Grant Number CNS-1836900. Any opinions, findings, and conclusions or recommendations expressed in this material are those of the authors and do not necessarily reflect the views of the National Science Foundation. The NASA University Leadership initiative (Grant \#80NSSC20M0163) provided funds to assist the authors with their research, but this article solely reflects the opinions and conclusions of its authors and not any NASA entity.\\
*The work was done while Alessandro Pinto was a Technical Fellow at the Raytheon Technologies Research Center. \\
This work was done as an outside activity and not in the author's capacity as an employee of the Jet Propulsion Laboratory, California Institute of Technology. } }

\author{\IEEEauthorblockN{Alessandro Pinto}
\IEEEauthorblockA{\textit{Raytheon Technologies*} \\
\textit{NASA Jet Propulsion Laboratory} \\
\textit{California Institute of Technology}\\
Pasadena, California \\
apinto@jpl.nasa.gov}
\and
\IEEEauthorblockN{Anthony Corso}
\IEEEauthorblockA{\textit{Aeronautics \& Astronautics Department} \\
\textit{Stanford University}\\
Stanford, California \\
acorso@stanford.edu}
\and
\IEEEauthorblockN{Edward Schmerling}
\IEEEauthorblockA{\textit{Aeronautics \& Astronautics Department} \\
\textit{Stanford University}\\
Stanford, California \\
schmrlng@stanford.edu}
}

\maketitle

\listoffixmes

\begin{abstract}
We apply a compositional formal modeling and verification method to an autonomous aircraft taxi system. We provide insights into the modeling approach and we identify several research areas where further development is needed. Specifically, we identify the following needs: (1) semantics of composition of viewpoints expressed in different specification languages, and tools to reason about heterogeneous declarative models; (2) libraries of formal models for autonomous systems to speed up modeling and enable efficient reasoning; (3) methods to lift verification results generated by automated reasoning tools to the specification level; (4) probabilistic contract frameworks to reason about imperfect implementations; (5) standard high-level functional architectures for autonomous systems; and (6) a theory of higher-order contracts. We believe that addressing these research needs, among others, could improve the adoption of formal methods in the design of autonomous systems including learning-enabled systems, and increase confidence in their safe operations.
\end{abstract}
    

\section{Introduction}
\label{sec:introduction}
Advances in Artificial Intelligence (AI) and Machine Learning (ML) are inspiring the research and development communities to implement systems that take on tasks traditionally under the control of human operators. The implementation of such systems, however, seems to be difficult to assure using traditional methods. It has been estimated\cite{kalraDrivingSafetyHow2016} that a self-driving car would need to be tested for tens of billions of miles to assess, with confidence, that the rate of fatal accidents is comparable to human-driven cars. For this reason, there is a strong interest in the use of \emph{formal methods} for the assurance of autonomous systems (see for example \cite{luckcuckFormalSpecificationVerification2020} for a recent survey). 

Of paramount importance is the use of formal specifications of requirements as they enable an assurance-driven design process. As stated in \cite{national2020advancing}, ``The unique nature of software essentially reduces the software safety problem to the safety of the software requirements provided to the programmers''. In fact, several previous works have developed approaches to requirement engineering for autonomous systems \cite{yahyaAutonomicComputingFramework2013,vassevAutonomyRequirementsEngineering2014,hobbsSystemsTheoreticProcess2022,pintoRequirementSpecificationAnalysis2021,vogelsangRequirementsEngineeringMachine2019}. Requirements focus on \emph{what a system, sub-system, or component is supposed to do, rather than on how it does it}, which suggests modeling the properties of its behaviors rather than the behaviors themselves. 

Requirements are also important in the development of complex systems that feature a tantamount complexity in their supply chain, operation, and ecosystem. The range of technologies involved in building an autonomous system may be too wide for a single company. They include sensors, actuators, mechanical systems, low-level controls, situational awareness, and several layers of decision-making. The envisioned Urban Air Mobility (UAM) system\cite{UrbanAirMobility2020}, for example, is composed of autonomous aircraft, fleet operators, service providers, supplemental data service providers, Unmanned Aircraft System (UAS) suppliers, the FAA, and the public. Each of these entities has a certain level of autonomy itself. The composition of all these services constitutes the UAM system. It is desirable to have well-defined requirements for each service to welcome different aircraft manufacturers or operators as they start their businesses. \emph{Compositional} design and verification methods\cite{benvenisteContractsSystemDesign2018} can be very effective in managing such complexity. Furthermore, autonomous systems evolve over time as a result of learning. Compositional methods can also enable incremental development and evolution.

In summary, we advocate for adopting a \emph{compositional formal framework for the definition of requirements for autonomous systems}. In particular, we focus on contract-based frameworks that have been studied in several previous works\cite{benvenisteContractsSystemDesign2018,Incer:EECS-2022-99} where contracts, operators and relations over contracts have been formalized. In \cite{bauerMovingSpecificationsContracts2012}, the authors show the connection between formal specification languages and assume-guarantee contracts. 
In this paper, we leverage this theoretical background work for the compositional modeling and verification of an autonomous taxi system for civil aircraft. Our  objective is not to develop a complete set of requirements, but to gain insights into the modeling process and to \emph{identify research directions to address some of the gaps we have encountered in our work}. From this standpoint, this study is broader in scope than the recently related approach presented in \cite{Cardoso2020TowardsCV}.

In our case study, an aircraft is given the goal of taxiing in an airport from a gate to the start of a runway. We have developed models along two different views: a protocol view that describes the possible sequences of events in the system, and a functional view that describes the properties of the data exchanged among components. Through this modeling effort, we have identified areas in which further research could have an impact on the adoption of formal methods and, as result, on our ability to gain confidence in the safety of autonomous systems which we briefly summarize here.  

\noindent \textbf{Heterogeneous contracts.} The decomposition of a model into different views offers a way to decompose the modeling and verification problem into sub-problems that can be efficiently addressed using different domain-specific tools. However, a formal approach to combining these views is lacking. \emph{If not addressed, this gap will leave room for informal statements regarding the properties of the combination of views}.

\noindent \textbf{Modeling productivity.} The development of formal models forces the elicitation of high-quality requirements and leads to proof of correctness, but it is a slow process that requires experts in formal methods. \emph{Increasing the model development productivity by subject matter experts will result in the wider adoption of formal methods and higher confidence in the safety of autonomous systems}.

\noindent \textbf{Explainability of formal methods.} When using tools such as model checkers or satisfiability solvers, a verification problem is encoded into the input language accepted by these tools. Verification results are obtained in a form that no longer relates to the original problem. Identifying the changes to be applied to a system resulting from the composition of contracts by interpreting verification results becomes impractical. \emph{This problem leads to a long and tedious model revision cycle that further reduces productivity.}

\noindent \textbf{Reasoning about approximate implementations.} Even if the requirements of a component can be formally stated, a perfect implementation may not be attainable -- as is the case for applications where data-driven approaches are used such as object recognition. Implementation uncertainty should be represented, and tools to reason about uncertainty in a compositional framework should be developed. \emph{Such tools could help assess operational risks, which is important for decision- and policy-makers}. 

\noindent \textbf{Standard robust architectures.} The requirement development process is iterative in nature. The efficiency gains realized through a compositional design and verification framework could become negligible if adding requirements led to radical changes in the functional architecture of the system. \emph{Standard functional architectures for autonomous systems robust to requirements evolution are essential to reduce development and verification time}. 

\noindent \textbf{Higher order contracts.} Autonomous operations often require the repeated use of different system capabilities a number of times which is context-dependent (e.g., moving and stopping at given points until a destination is reached). Proving mission-level properties starting from contracts of these capabilities requires methods that go beyond traditional compositional reasoning. \emph{Addressing this gap enables proving system-level properties in general contexts}.

\section{Description of the taxi scenario}
\label{sec:taxi-scenario}

\begin{figure}
    \begin{center}
        \includegraphics[width=0.95\columnwidth]{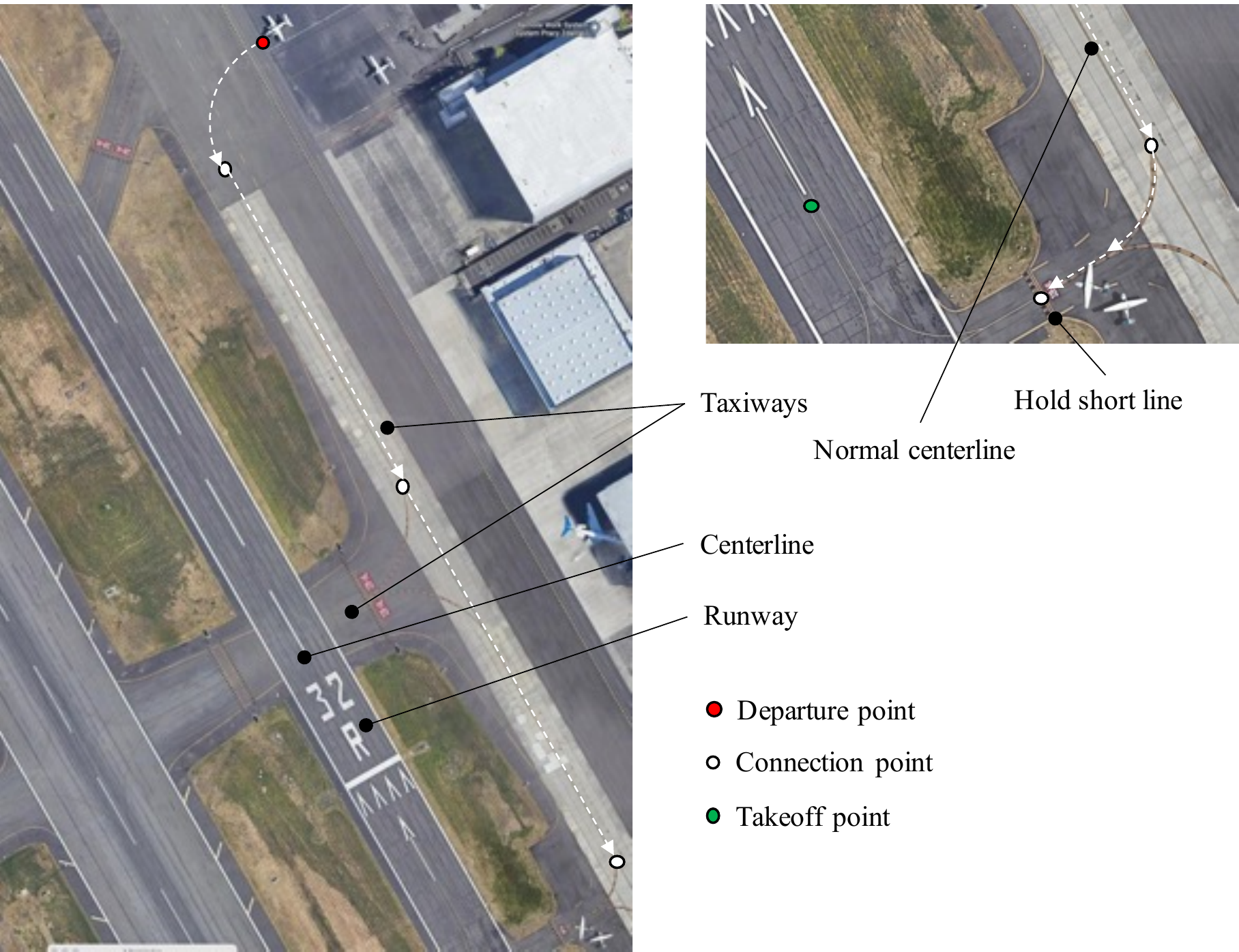}
        \caption{The taxi scenario. \label{fig:scenario}}
    \end{center}
\end{figure}

The elements of a typical taxi operation are shown in Figure \ref{fig:scenario}. A general aviation aircraft is at its initial departure point (marked as a red dot) at rest. The aircraft accepts a takeoff command indicating a takeoff point (marked as a green dot). When the command is received, the aircraft is expected to follow a continuous trajectory (marked as a white dashed line) over the centerline of the taxiways that reaches the takeoff location and that goes through some intermediate points (marked as white dots). The aircraft is supposed to stop at intersections marked with a hold short sign on the ground perpendicular to the taxiway. 
The concept of operation for this scenario includes an initial phase where the aircraft receives a detailed map of the airport, its current conditions, weather reports, and any other relevant information that may affect the flight. We assume that the airport is used by both human pilots and autonomous aircraft and that changes to the infrastructure are not acceptable. Thus, the aircraft must be able to perceive the environment using location information from standard sensors such as a GPS, or by observing airport signage. In particular, the aircraft shall be able to move along centerline markings and to identify and stop at a safe distance from a hold-short line. 

\begin{figure}
    \begin{center}
        \includegraphics[width=0.8\columnwidth]{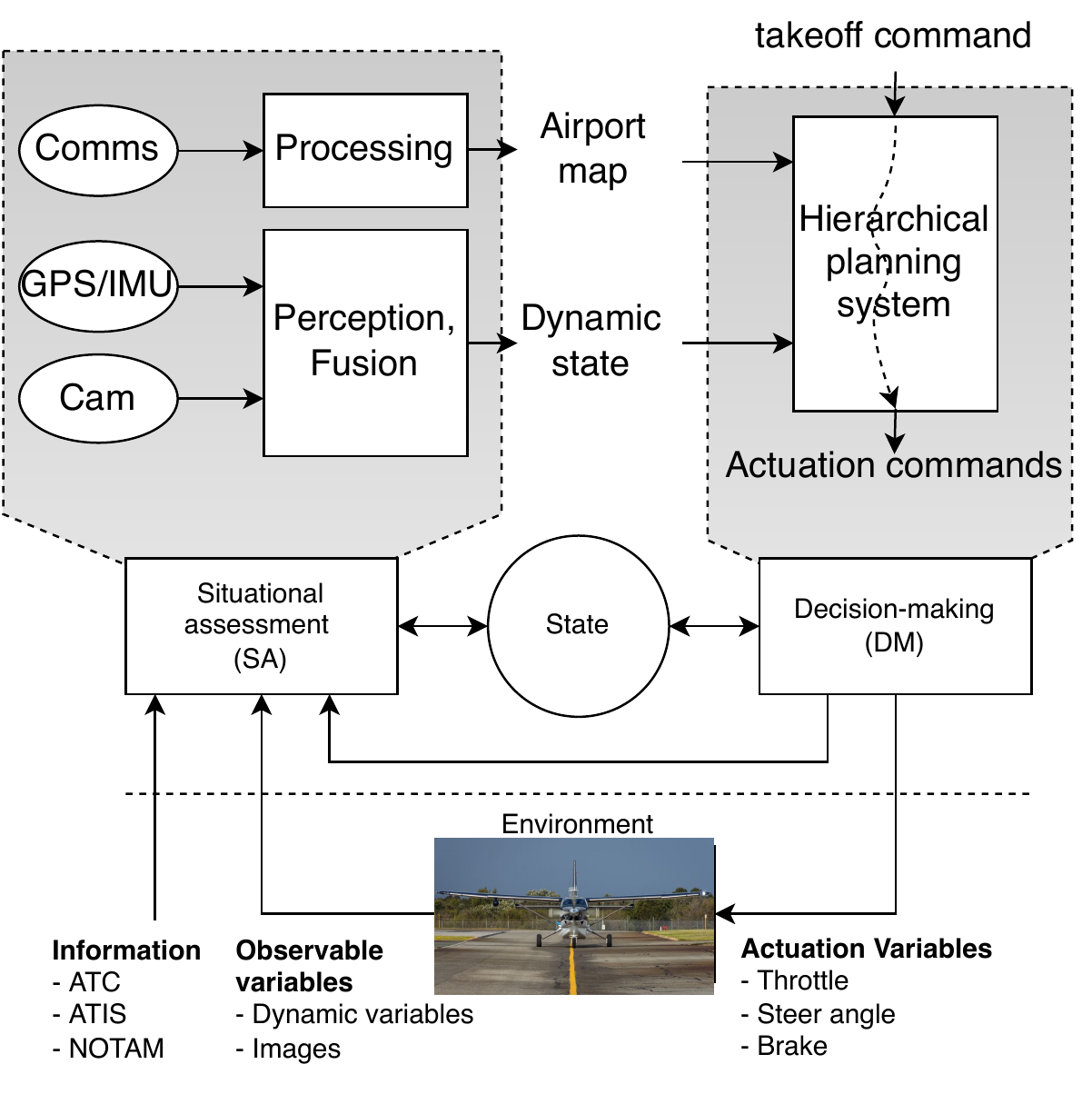}
        \caption[High-level architecture]{High-level architecture of the system that plans for, and executes a taxi command. \label{fig:HLA}}
    \end{center}
\end{figure}
A high-level functional architecture of the system that controls the aircraft is shown in Figure \ref{fig:HLA}.  
\emph{Observable} variables are measured directly from the environment. In our case, they include the dynamic state of the aircraft and images from a camera. \emph{Controllable} variables influence directly the environment. In our case, they include throttle, steering angle, and brake commands. The function of the situational assessment sub-system is to maintain an estimate of the state of the system, while the function of the decision-making sub-system is to determine the value of the controllable variables at each point in time. These values (also referred to as decisions) must be selected in such a way as to achieve the goals of the system, or to maximize a reward function, while satisfying several constraints, such as safety. 

The top part of Figure \ref{fig:HLA} shows some details of the two main functions for our scenario. The SA sub-system includes sensors such as a GPS, an IMU, and a camera. 
A camera might be needed to improve control accuracy using a 
visual servoing approach.
In particular, images from the vision sensor could be processed to estimate position and orientation relative to the center line, and distance from a hold-short line. Information from these sensors is processed and fused together to assess the dynamic state of the aircraft. The state is part of a world model which includes a map of the airport where the aircraft is located (gathered by communicating with several services such as Air Traffic Control (ATC), and Notice to Airman (NOTAM)), and the health status of the aircraft subsystems. 

\subsection{Modeling methodology}
In this article we use a modeling methodology inspired by \cite{pintoModelingMethodologyAutonomous2022}.  
We start by defining \emph{a set of decision-making (DM) capabilities} that enable the aircraft to taxi autonomously. The definition of DM capabilities arises from the decomposition of the DM problem into layers of decisions. This decomposition is necessary to limit the complexity of the DM problem. In our case, we will use a traditional autonomy stack which includes  
route generation, trajectory generation, and trajectory following. The specification of these capabilities is modeled by the assumptions they make on the goals assigned to them and the behavior of their environment, while the guarantees are properties of the generated decisions. In order to make decisions, these capabilities require an estimate of the current state at the level of abstraction where the decisions are made. For example, route generation requires an estimate of the current node in the roadmap where the aircraft is. Finally, the accuracy of these estimates impacts the feasibility and accuracy of the decisions made by the DM capabilities. Thus, the result of the design of the decision-making sub-system also provides a specification for the situational assessment sub-system and a set of assumptions that define the envelope where the system can operate.

We will often refer to two views of the world: internal and external. The internal world model is an estimate of the external world. The estimate is generated by the autonomous system through its situational assessment sub-system. The variables in the internal world model will have the same names as the corresponding variables in the external world models, but they will be marked by an accent. For example, the current coordinate of the aircraft will be denoted by $x$ and $y$, while their estimates in the internal world model will be denoted by $\hat{x}$ and $\hat{y}$. 

The decision-making capabilities can only access the internal world model. However, safety and performance requirements are expressed in terms of the external world model. The discrepancy between the two represents the ignorance of the autonomous system about the world and needs to be bounded in order to guarantee safety. The relation between the internal and external world models is defined by the situational awareness function and by the actuation system. 

Modeling DM capabilities and SA functions requires a formal treatment of assumptions and guarantees. We leverage the contract-based framework \cite{bauerMovingSpecificationsContracts2012} for this purpose. In particular, we will use several operators that allow us to partition the specification into different scenarios, analysis domains, and abstraction layers. 

\section{Background}
\label{sec:background}

\subsection{Contracts}
We will use a contract-based approach \cite{benvenisteContractsSystemDesign2018,Incer:EECS-2022-99} to develop our compositional framework,
and we will adopt the formalism in \cite{bauerMovingSpecificationsContracts2012} which we briefly summarize here (please refer to \cite{bauerMovingSpecificationsContracts2012} for details). A specification theory is a triple $(\mathcal S, \otimes, \leq)$, where $\mathcal S$ is a set of specifications, $\otimes$ is a binary operation over specifications called parallel composition, and $\leq$ is a reflexive, transitive, and compositional relation over specifications called \emph{refinement}. Given two specifications $S$ and $T$, $S\wedge T$ is the most general specification that refines both $S$ and $T$, while $S\vee T$ is the more specific specification that is refinement by both $S$ and $T$. 
The relativized refinement relation $\leq_E$, instead, is defined as follows: given three specifications $S$, $E$, and $T$, $S \leq_E T$ if and only if for all
$E' \leq E$, $S \otimes E' \leq T \otimes E'$. In words, $S$ refines $T$ when both are considered in an environment $E'$ that refines $E$. 

A contract theory is built over a specification theory. Specifically, a contract is a
pair $C = ( A, G)$ where $A$ and $G$ are specifications called assumption and guarantee,
respectively. The environment semantics of a contract is the set of specifications that
refines the assumption: 
$\semantics{C}_{env} = \{E \in \mathcal S | E \leq A\}$. 
The implementation semantics
of a contract is the set of specifications that satisfy the guarantee $G$ under the assumption
$A$: 
$\semantics{C}_{impl} = \{I \in \mathcal S | I \leq_A G\}$.
Two contracts are semantically equivalent if their
environment and implementation semantics are the same, respectively. Moreover, a contract $C$ is in normal form if for all specification $I \in \mathcal S$, $I \leq_A G$ if and only if $I \leq G$, which also means that for a contract $C^{nf} = ( A^{nf} ,G^{nf} )$ in normal form,
$\semantics{C^{nf}}_{impl} = \{I \in \mathcal S | I \leq G\}$.
In some cases, depending on the specification
theory, a contract $C = ( A, G)$ can be transformed into a semantically equivalent
contract $C^{nf} = ( A, G^{nf} )$ in normal form by weakening its guarantee. Under some assumptions on the specification theory, the normal form of the guarantee of a
contract can be computed as $G^{nf} = G \wedge \neg A$ (where $\neg A$ is the set of environments that do not
refine $A$). In this case, a contract in normal form is also called saturated.

Contracts are related by a refinement relation $\preceq$. A contract $C' = ( A' , G')$ refines
$C = ( A, G)$ if and only if $\semantics{C'}_{env} \subseteq \semantics{C}_{env}$, and $\semantics{C'}_{impl} \subseteq \semantics{C}_{impl}$.
It can be
shown that this condition corresponds to $A \leq A'$ and $G' \leq_{A} G$, and if contracts are
in normal form, then the latter can be written as $G' \leq G$.

In this paper, we are interested in the verification of correct refinements. If $C$ is a specification contract, then the set $\{C_1, \ldots , C_n\}$ is a correct decomposition of $C$ if and only if the following conditions hold \cite{leContractbasedRequirementModularization2016}:
\begin{equation}
\begin{split}
    \bigwedge_{1 \leq i \leq n} G_i^{nf} & \leq G^{nf} \\
    A \wedge \bigwedge_{1 \leq j \neq i \leq n} G_j^{nf} &\leq A_i, \qquad \forall i \in [1,n]
\end{split}
\label{eq:contract-refinement-conditions}
\end{equation}

Finally, we will leverage the conjunction operator for contracts \cite{Incer:EECS-2022-99}. Let $C_1 = ( A_1 , G_1^{nf} )$ and
$C_2 = ( A_2 , G_2^{nf} )$ be two contracts, then $C_1 \wedge C_2 = (A_1 \vee A_2, G_1^{nf} \wedge G_2^{nf})$ is their conjunction. The conjunction operator allows modeling and analyzing systems along different views. Let $C=C_f\wedge C_t$, $C'=C_f'\wedge C_t'$, and $C''=C_f''\wedge C_t''$ be contract decomposed along the $f$unctional and $t$iming views. Then, it can be shown that if $\{C_f',C_f''\} \leq C_f$ and $\{C_t',C_t''\} \leq C_t$, then $\{C',C''\} \leq C$.

\subsection{Specification theories}
\label{ssec:specification-theories}
In this paper, we will use mainly two specification theories: Many-Sorted First Order Logic (MS-FOL) and Linear Temporal Logic with a Past modality\cite{markeyTemporalLogicExponentially2003} (PLTL), which we briefly review in this section. Aside from these illustrative choices, we note that there are many other possible specification theories that can be considered, such as probabilistic variants of FOL or PLTL, and that can be used to develop contract theories.

A MS-FOL signature includes a set of type symbols $\types$, a set of  constant symbols $\constants$, a set of predicate symbols $\predicates$, a set of function symbols $\functions$, and a countably infinite set of variable symbols $\variables$. Constants and variables have a type. A constant $c \in \constants$ of type $T \in \types$ is denoted by $c:T$. Similarly, a variable $v \in \variables$ of type $T$ is denoted by $v:T$. Constants and variables are terms in the MS-FOL language. A function symbol $f \in \functions$ has a signature $(T_1,\ldots,T_n) \rightarrow T_0$ where $T_i \in \types$ are types. A function is a term $f(t_1, \ldots, t_n) : T_0$ where $t_i : T_i$ are terms, and has type $T_0$. A predicate symbol $p \in \predicates$ also has a signature $(T_1,\ldots,T_m)$ where $T_i \in \types$. A predicate is an atomic formula $p(t_1, \ldots, t_n)$ where $t_i : T_i$ are terms. Given formulas $\phi$ and $\psi$, and variable $v \in \variables$ of type $T \in \types$, $\neg \phi$, $\phi \wedge \psi$, and $\forall v:T. \phi$ are also formulas. 
The semantics of this language is defined in terms of interpretations. An interpretation is a pair $\interpretation=(D,\mu)$ where $D:\types \rightarrow 2^\objects$ maps each type symbol to a set of objects, and $\mu$ maps symbols to their interpretations. In particular: 
for a constant $c:T$, $\mu(c) \in D(T)$ is an object; 
for a variable $v:T$, $\mu(v) \in D(T)$ is an object; 
for a function symbol $f$ with signature $(T_1,\ldots,T_n) \rightarrow T_0$, $\mu(f)$  is the definition of a map from $D(T_1) \times \ldots \times D(T_n)$ to $D(T_0)$; 
and for a predicate symbol $p$ with signature $(T_1,\ldots,T_m)$, $\mu(p) \subseteq D(T_1) \times \ldots \times D(T_n)$ is a relation. 
We also extend $\mu$ to functions: $\mu(f(t_1,\ldots,t_n)) = \mu(f)(\mu(t_1),\ldots,\mu(t_n))$.
Entailment is a relation between interpretation and formulas defined recursively as follows: $\interpretation \models p(t_1,\ldots,t_n) \iff (\mu(t_1),\ldots,\mu(t_n)) \in \mu(p)$,  $\interpretation \models \neg \phi \iff \interpretation \neg\models phi$, $\interpretation \models \phi \wedge \psi \iff \interpretation \models \phi$ and $\interpretation \models \psi$, $\interpretation \models \forall v:T.\phi \iff \interpretation \models \phi(v\leftarrow o)$ for all $o \in D(T)$.  

We now review the syntax and semantics of PLTL. Let $Prop$ be a finite set of propositions. A PLTL formula is defined as follows: $\phi:=\neg\phi|\phi\wedge \psi|\phi \ltlU \psi|\psi \ltlS \phi|\ltlX \phi | \ltlXm \phi | p$, where $\phi$ and $\psi$ are formulas and $p \in Prop$. The temporal modalities can be interpreted in plain English as ``until''($\ltlU$), ``since''($\ltlS$), ``next''($\ltlX$),  ``previously''($\ltlXm$), and ``releases'' ($\ltlR$). The semantics of a formula is defined over infinite sequences of sets in $2^{Prop}$. Specifically, given a sequence $\pi$ and an index denoting the position in the sequence:
\begin{equation*}
    \begin{split}
        \pi,i &\models \phi \ltlU \psi \ iff \ \exists k\geq i. (\pi,k\models \psi \wedge \forall i \leq j < k. \pi,j\models \phi) \\ 
        \pi,i &\models \ltlX \phi \ iff \ \pi,i+1\models \phi \\
        \pi,i &\models \phi \ltlS \psi \ iff \ \exists k\leq i. (\pi,k\models \psi \wedge \forall k < j \leq i. \pi,j\models \psi) \\ 
        \pi,i &\models \ltlXm \phi \ iff \ \pi,i-1\models \phi \\
        \pi,i &\models \phi \ltlR \psi \ iff \  \forall k \geq i. (\pi,k\models \psi \vee \exists i \leq j < k. \pi,j\models \phi)
    \end{split}
\end{equation*}
The logical connectives $\neg$ and $\wedge$ have the usual semantics. Given these modalities, it is possible to define others as usual. In particular, $\ltlF \phi$ means that eventually $\phi$ holds, and $\ltlFm \phi$ means that in the past $\phi$ held. 

Technically speaking, a specification in MS-FOL is a pair $(\Sigma,\phi_\Sigma)$ where $\Sigma$ is a MS-FOL signature, and $\phi$ is a MS-FOL formula, and a specification in PLTL is $(Prop,\phi_{Prop})$ where $Prop$ is a set of atomic propositions and $\phi$ is a PLTL formula. However, we will consider a universal FOL signature, and a universal set of propositions so that our specifications are going to be just formulas.

\section{Compositional modeling and verification of the taxi scenario}
\label{sec:compositional-modeling}

In this section, we use contracts to develop a model of a classical layered architecture for a nominal autonomous taxi scenario. Our goal is not to develop a complete model but rather to identify gaps in modeling and verification that need to be addressed through further research.

\begin{figure}
\centering
\includegraphics[width=0.9\columnwidth]{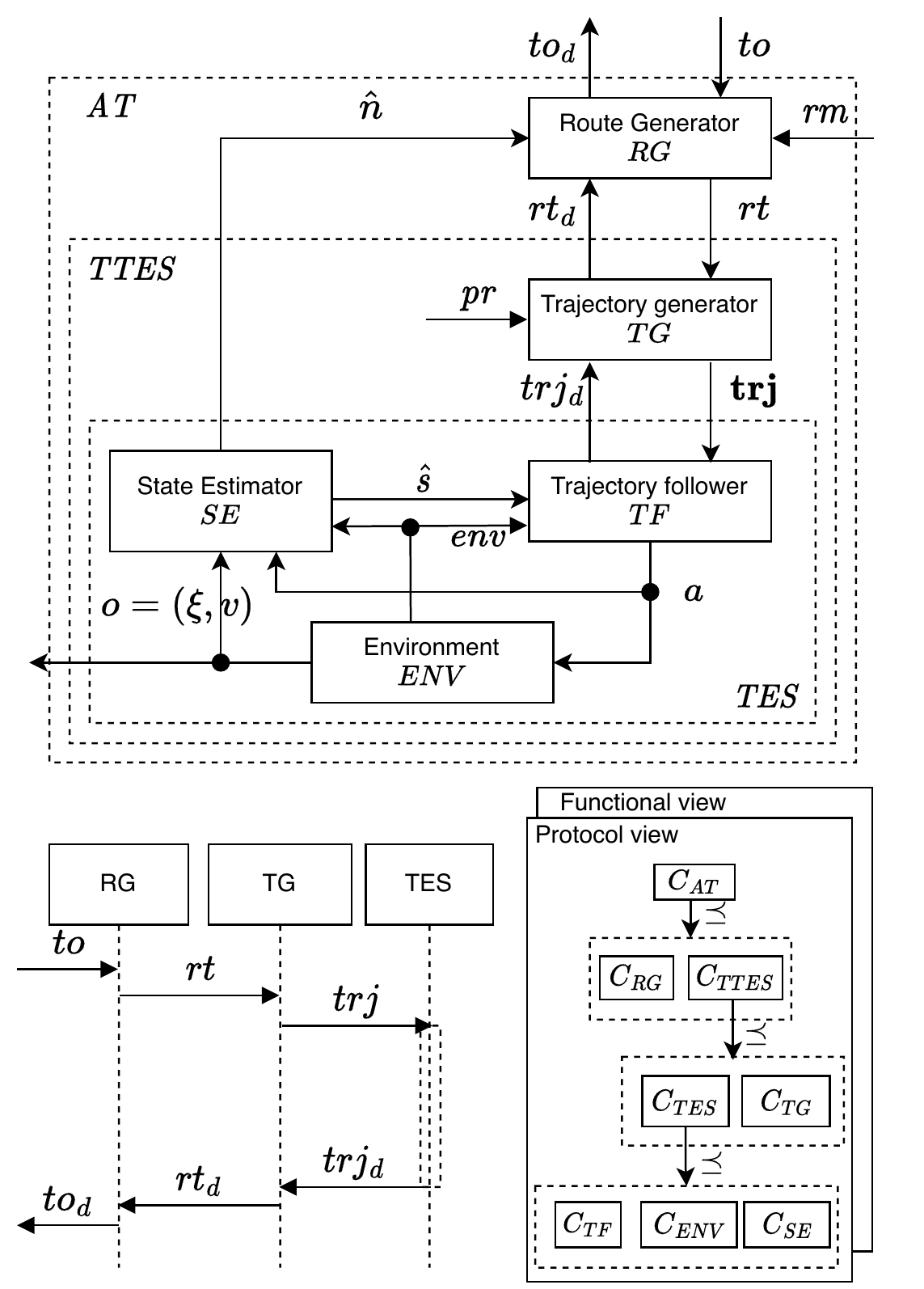}
\caption{Diagram of the multi-layer architecture for the autonomous taxi capability. \label{fig:first-level}}
\end{figure}

The top part of Figure \ref{fig:first-level} shows a hierarchical structure with several components. The route generator $\RG$ receives a takeoff command as a takeoff node $\vto$, and the estimate of the initial node $\hat{n}$ over a roadmap $\vrm$ which we assume to be accurate and known to $\RG$. The route $\vrt$ is then sent to a trajectory generator $\TG$ which generates a trajectory $\vtrj$. The trajectory is finally sent to a trajectory follower $\TF$ connected in closed loop with the environment $\ENV$ and a state estimator $\SE$. The trajectory follower decides the values of the actuation variables $a=(\vth,\vst,\vbr)$ (denoting throttle, steering, and breaking) over time. Once the trajectory has been fully executed, $\TF$ informs $\TG$ with a binary signal $\vtrj_d$ (trajectory done). In turn, $\TG$ informs $\RG$ with a binary signal $\vrt_d$ (route done). Finally, $\RG$ informs the external entity that sent the command with a binary signal $\vto_d$ (takeoff command done). 
Notice that in the nominal case 
a component at a given layer assumes that the goals given to the next layer will eventually be achieved. As shown in the bottom right portion of Figure~\ref{fig:first-level}, modeling and verification proceed in successive refinement steps. First, it should be shown that $\{C_{\RG},C_{\TTES}\} \preceq C_{\AT}$, then that $\{C_{\TES},C_{\TG}\} \preceq C_{\TTES}$, and finally that $\{C_{\TF},C_{\ENV},C_{\SE}\} \preceq C_{\TES}$. To manage the complexity of these verification problems, we decompose the model into two views as described next\footnote{Detailed models for this example can be found at \url{https://github.com/alessandro-pinto/icaa2023models}}.

\subsection{Heterogeneous views}
\label{sec:Heterogeneous-contracts}
We start by modeling the two components $\RG$ and $\TTES$. The specification language needs to be expressive enough to capture the causality of events and the properties of data exchanged by these components. There is a trade-off between the expressive power of formal languages and the complexity of the verification problem. For example, OCRA~\cite{cimattiOCRAToolChecking2013} is a contract-specification and refinement analysis tool based on the HRELTL~\cite{cimattiRequirementsValidationHybrid2009} language which could  be used for our purposes. These languages provide support for temporal properties, for standard data types such as Boolean, integer, and enumeration, and arrays of elements of such types. With this limited set of types, modelers have to spend a considerable amount of time in the model encoding process as we will see later. Most importantly, the entailment problem for such languages is often undecidable. 

Instead, we can leverage the compositional framework of contracts to split the model into separate views. For example, a ``protocol'' view could be used to capture the causality of events for each component, while a ``functional'' view could be used to capture the data properties that are supposed to be invariant. Different views can, in principle, be modeled using different languages, and contract refinement can be verified using different verification tools.

The ``protocol'' view can be captured well using LTL formulas. For example, the formula $\phi_{er} = (r \Rightarrow \ltlXm(\neg r \ltlS e))$ means that a response $r$ is generated only for an event $e$. Let $\phi_{er}(e',r')$ be the same formula where $e$ has been renamed to $e'$ and $r$ to $r'$. The contract for the route generator protocol view can be expressed as follows:

\begin{subequations}
    \begin{align}
    A_{\RG}^{(p)} &= \ltlG(\vto \Rightarrow \ltlX( \neg \vto \ltlU \vto_d)) \wedge  \label{eq:a-rg-p-1} \\    
                &\wedge  \ltlG(\vrt \Rightarrow \ltlF \vrt_d) \wedge \phi_{er}(\vrt,\vrt_d) \label{eq:a-rg-p-2} \\ 
    G_{RG}^{(p)} &= \ltlG(\vto \Rightarrow \ltlF \vrt) \wedge \phi_{er}(\vto,\vrt)  \label{eq:g-rg-p-1} \\
     &\wedge \ltlG(\vrt_d \Rightarrow \ltlF \vto_d) \wedge \phi_{er}(\vrt_d,\vto_d)  \label{eq:g-rg-p-2}
    \end{align}
\end{subequations}

The component assumes that: (1) after the environment sends a $to$ command, no new $\vto$ commands are sent to $\RG$ until the previous one has been completed (\ref{eq:a-rg-p-1}); (2) the $\TTES$ component eventually responds to an $\vrt$ command with $\vrt_d$, and the $\TTES$ component does not issue $\vrt_d$ unless there has been an $\vrt$ command in the past (\ref{eq:a-rg-p-2}). On the other hand, $\RG$ guarantees that: (1) $\vto_d$ is issued after receiving (and only after receiving) $\vrt_d$ (\ref{eq:g-rg-p-1}); (2) $\vrt$ is issued after receiving (and only after receiving) $\vto$ (\ref{eq:g-rg-p-2}). The $\TTES$ component does not make any assumption on the environment: $A_{\TTES}^{(p)} = \top$. In order words, it always accepts new routes. It guarantees to respond only to route requests: $G_{\TTES}^{(p)} = \ltlG(\vrt \Rightarrow \ltlF \vrt_d) \wedge \phi_{er}(\vrt,\vrt_d)$. Finally, the requirements that the composition must satisfy can also be modeled by a specification contract where $A_{\AT}^{(p)} = \ltlG(\vto \Rightarrow \ltlX(\neg \vto \ltlU \vto_d) )$, and  $G_{\AT}^{(p)} =  \ltlG(\vto \Rightarrow \ltlF \vto_d)$.
The contracts above can be modeled in OCRA which can automatically check if $\{C_{\RG}^{(p)}, C_{\TTES}^{(p)}\} \preceq C_{\AT}^{p}$. \footnote{The model can be found at \url{https://github.com/alessandro-pinto/icaa2023models/blob/main/AT/protocol/AT.oss}}.

The functional view captures the properties that input and output data must satisfy. The $\RG$ component makes several assumptions on its inputs including the well-formedness of the roadmap $\vrm$, the membership of the initial node $\hat{n}$ and node $to$ to the roadmap, and the existence of a path that connects the two. It guarantees to find a valid path in the roadmap that starts at $\hat{n}$ and ends in $to$. To arrive at an MS-FOL model for this view we start with some formal definitions for roadmaps, routes, and trajectories. A roadmap $\vrm=(N,L)$ is a graph where $N$ is a set of nodes and $L \subseteq N \times N$ is a set of links. Functions $\vloc:N \rightarrow \mathbb{R}^2$ and $\vlbl:N \rightarrow \{g,t,c,h\}$ associate coordinates and labels to nodes, respectively, where $g$ stands for gate, $t$ for takeoff, $c$ for connection point, and $h$ for hold short. Links are annotated with a length $\vlen:L\rightarrow \mathbb{R}_{\geq 0}$. A route of length $k$ in the roadmap is  a sequence of links $(l_1,...,l_k)$, all belonging to $L$, where for all $i\in [1,k-1]$, $\vdst(l_i)=\vsrc(l_{i+1})$. Given two nodes $n_1$ and $n_2$ in the roadmap, $\vrts(rm,n_1,n_2)$ is the set of routes from $n_1$ to $n_2$.  Finally, we need several definitions for trajectories. Each link $l$ in the roadmap is associated with a continuous motion primitive $\vpr(l):[0,\vlen(l)] \rightarrow \mathbb{R}^2\times [0,2\pi)$ which maps the linear distance from the start of the motion primitive to a pose. This function tracks the centerline of the taxiway that the link $l$ represents. 
Given a route $\vrt=(l_1...,l_k)$, $\vprs(rt)$ is the concatenation of $(\vpr(l_1),...,\vpr(l_k))$ which, by definition of $\vpr$, is a continuous trajectory of total length $|\vprs(rt)| = \vlen(l_1) + \ldots + \vlen(l_k)$. Also,  $\vtrjs(rm,n_1,n_2) = \{\vprs(rt)| \vrt \in \vrts(n,n',rm) \}$ is the set of trajectories from $n_1$ to $n_2$ in the roadmap. Finally, a distance function $d(\vrt,i)$  maps a route and an integer $i \in [1,k]$ to the total distance from the first node to the destination of the $i$-th link. This function is recursively defined as follows: $d(\vrt,0) = 0$ and $d(\vrt,i) = d(\vrt,i-1) + \vlen(l_i)$ for $i \in [1,k]$. 

The autonomous taxi system has three top-level requirements: it shall follow the centerline, shall stop short of the hold-short lines, and shall stay below a prescribed maximum velocity and acceleration at all times. Consider the first of these requirements ($cl$) formalized as the following guarantee:
\begin{equation}
    \begin{split}
        G_{\AT}^{(cl)} &= \exists \vtrj \in \vtrjs(\vrm,\hat{n},\vto). \\
        &\qquad \forall p \in [0,|\vtrj|]. \\
        &\quad\qquad \exists t_p \in \mathbb{R}_{\geq 0}. \\
        &\quad\quad\qquad p=\int_{t_0}^{t_p} v(t) dt \wedge \xi(t_p) \in \mathcal B_\delta (\vtrj(p))
    \end{split}
\end{equation}
This requirement essentially means that the aircraft follows a valid trajectory within some allowed error defined by a ball of radius $\delta$. In this definition, $t_0$ is the time when the $\vto$ command is sent to the system, and $v(t)$ is the velocity of the aircraft at time $t$. The composition of the $cl$ views of $\RG$ and $\TTES$ must satisfy this requirement.

In order to use automated reasoning tools, we need to encode these mathematical definitions using an appropriate specification language such as MS-FOL which is well-supported by efficient theorem provers such as z3\footnote{\url{https://github.com/Z3Prover/z3}} and cvc5\footnote{\url{https://cvc5.github.io}}. We will focus more on the description of the encoding process for the most important definitions. We define $Node$s, $Link$s, $Roadmap$s, $Route$s, and trajectories ($Trj$) as types. Labeling functions are encoded as functions. For example, the source function is defined as $\vsrc:Link \rightarrow Node$. To model membership of nodes and links to roadmaps, we use two predicates $InN (Node,Roadmap)$, and $InL(Link,Roadmap)$. $linkAt: (Route,Int) \rightarrow Link$ is a function that maps a route and an index to the corresponding link in the route. Moreover, $length:Route \rightarrow Int$ denotes the number of links in the route, such that given a route $\vrt$, $linkAt(\vrt,i)$ is meaningless for $i$ outside the range $[1,length(\vrt)]$. We also define a predicate $InBall(Pose,Pose,\mathbb R_{\geq 0})$ with the intended meaning that $InBall(p_1,p_2,\delta)$ is true if and only if $p_1$ is within a ball or radius $\delta$ from $p_2$. Finally, the relation between linear distance and time is simply modeled by a function $intg:\mathbb{R}_{\geq 0} \rightarrow \mathbb{R}_{\geq 0}$, such that $intg(t)$ is the distance from the beginning of a trajectory until time $t$. Clearly, all these functions have properties that should also be encoded as background axioms in FOL. However, this is needed only if the solution to our verification problem depends on such properties, and this is not necessarily the case. In fact, when checking refinement using the conditions in Equation \ref{eq:contract-refinement-conditions}, it might be enough to show that the assumptions of a component are directly supported by the guarantees of another, and the proof may not depend on background axioms. In our case, however, it is important to reason about neighborhoods of different sizes when proving refinement. Thus, we would need to add at least the following axiom to our background knowledge: $\forall \delta_1,\delta_2,p_1,p_2. \delta_1 \leq \delta_2 \Rightarrow (InBall(p_1,p_2,\delta_1) \Rightarrow InBall(p_1,p_2,\delta_2))$. Moreover, we would also need to say that ``being in a neighborhood'' is a symmetric property: $\forall \delta,p_1,p_2. (InBall(p_1,p_2,\delta) \Rightarrow InBall(p_2,p_1,\delta))$ 
Finally, we use a predicate $ExistRt(Roadmap,Node,Node)$ such that $ExistRt(\vrm,n_1,n_2)$ is true if and only if  $\vrts(\vrm,n_1,n_2)$ is not empty. Informally, the $\RG$ component assumes that the roadmap $is$ $valid$, $\vto \in N$, $\hat{n} \in N$ and $\vrts(rm,\hat{n},to)$ is not empty. It guarantees to find a valid route $\vrt \in \vrts(rm,\hat{n},to)$\footnote{We are not requiring $RG$ to find the best route, but just one.}. 
Formally, we can state assumptions and guarantees of $\RG$ as follows:

\begin{equation*}
\begin{split}
A_{RG}^{(cl)} &= ValidRm(\vrm) \wedge InN(\hat{n},\vrm) \wedge InN(\vto,\vrm) \wedge \\
       &\wedge ExistRt(\vrm,\hat{n},\vto) 
       \end{split}
\end{equation*}
\begin{equation*}
\begin{split}
G_{RG}^{(cl)} &= src(linkAt(\vrt,1)) = \hat{n} \wedge \\
       &\wedge dst(linkAt(\vrt,length(\vrt))) = to \wedge \\
       &\wedge \forall i \in [1,length(\vrt)-1] . \\
       &\qquad dst(linkAt(\vrt,i)) = src(linkAt(\vrt,i+1)) \wedge \\
       &\wedge \forall i \in [1,length(\vrt)] . InL(linkAt(\vrt,i),rm)
\end{split}
\end{equation*}

We now move to define the contract for $\TTES$. 

\begin{equation}\label{eq:ttes-contract}
\begin{split}
A_{TTES}^{(cl)} &= src(linkAt(\vrt,1)) = \hat{n} \wedge \\
        & \forall i \in [1,length(\vrt)-1] . \\
        &\qquad dst(linkAt(rt,i)) = src(linkAt(rt,i+1)) \\
G_{TTES}^{(cl)} &= \forall i\in [1,length(\vrt)]. \forall p \in [d(\vrt,i-1),d(\vrt,i)]. \\
         &\quad \exists t.\ p=intg(t) \wedge \\
         &\wedge InBall(\xi(t),pr(linkAt(\vrt,i),p-d(i)),\delta) 
\end{split}
\end{equation}

The $\TTES$ component assumes that $\vrt$ is a valid route, and controls the system so that the observed variable $\xi$ follows the centerline. The guarantee could also include velocity constraints as $(lbl(dst(linkAt(rt,i))) = h \Rightarrow \exists p' \in [d(rt,i-1),d(rt,i)-\beta], t'. p'=intg(t') \wedge v(t') = 0)$, meaning that the aircraft stops short of a node marked as $h$.
We can now use an SMT solver to check the refinement between the specification and the composition of $RG$ and $TTES$\footnote{The model can be found at \url{https://github.com/alessandro-pinto/icaa2023models/blob/main/AT/function/AT.smt2}}. 

\paragraph{Gap analysis and research directions}
\label{par:heterogeneous-contracts-lessons} 
This initial system model exposes several gaps in the compositional formal specification and verification of autonomous systems. The modeling effort, even for a simplified system, is generally high. This problem can be traced back to two gaps. First, the \emph{lack of a language at a higher level of abstraction} than languages (such as Othello\cite{cimattiOCRAToolChecking2013} or smtlib2\cite{BarFT-RR-17}) used as inputs to solvers. Secondly, the \emph{lack of libraries of formal models for common concepts such as graphs, paths, and trajectories}, which are used in autonomous systems that move in a physical space, increases the modeling time. Research is needed in the development of a specification language and a set of libraries to speed up modeling. The definition of such language should keep in mind the usual trade-off between expressive power and decidability. Moreover, the translation of models from this language into standard inputs to lower-level solvers should be doable with minimal translation efforts. Libraries should also be carefully engineered to enable efficient verification since the run-time performance of common solvers is sensitive to the structure of the input model. Finding suitable abstractions for common concepts such as the ones presented in this section is key to reducing the complexity of the verification problem.  

Finally, while compositional methods allow the use of several views which can be modeled and verified with different tools, there is a \emph{lack of methodologies and theoretical results to manage the interaction among these heterogeneous sets of tools}. We have not formally combined the analyses from the protocol and functional views into one single contract, but rather left implicit that the variables used in different views are disjoint, and the models are constructed in a way that takes care of their dependencies. Research is needed in the definition of computational procedures to combine different solvers to verify models where there is an interaction among views.

\subsection{Parametric contracts and control abstraction.} 
\label{sec:recursive-contracts-controls}
We now decompose $TTES$ into sub-components while using $C_{\TTES}$ as specification. Component $\TTES$ is first decomposed into a trajectory generator $\TG$ and the $\TES$ subsystem. We consider a $\TG$ component that generates several trajectory segments instead of one. Each segment starts at the gate or at a hold short location and ends in a hold short location or at the takeoff location, without any other hold short location in between. For each segment, $\TG$ generates a pose trajectory and a velocity profile that starts at zero, reaches a maximum speed if possible, and then decreases to zero at the end. 

We first define the protocol view of these components. The input to $\TTES$ is not only a route event $\vrt$, but also an integer number $n_h$ which represents the number of hold-short locations in the route. If we assume that a route starts at a gate and ends at a takeoff location, then the number of segments that need to be generated is $n_h+1$. For each segment, $\TG$ must generate a trajectory event $\vtrj$ and wait for a response $\vtrj_d$ indicating that the trajectory segment has been executed. Thus, we have a protocol view with an integer parameter $n_h$, whose contract depends on the parameter. For $n_h=0$, there will be only one $\vtrj$ event and one $\vtrj_d$ response. For $n_h=1$ there will be one $\vtrj$ event followed by a $\vtrj_d$ response, followed by another $\vtrj$ event and a corresponding $\vtrj_d$ response, and so on. The objective is to define such a parametric contract whose expression depends on $n_h$. 

To address this problem, we can proceed in two ways. We can assume that there is an upper bound on $n_h$ called $\overline{n}_h$ and prove refinement against the specification of $\TTES$ for all values of $n_h \in [0,\overline{n}_h]$. This might be feasible if the upper bound is known and small. The other option is to use an inductive argument and prove that: (1) refinement holds for $n_h=0$ and 
 (2) assuming that it holds for $n_h \geq 0$, then it holds for $n_h+1$. 
 
\begin{figure}
    \centering
    \includegraphics[width=0.95\columnwidth]{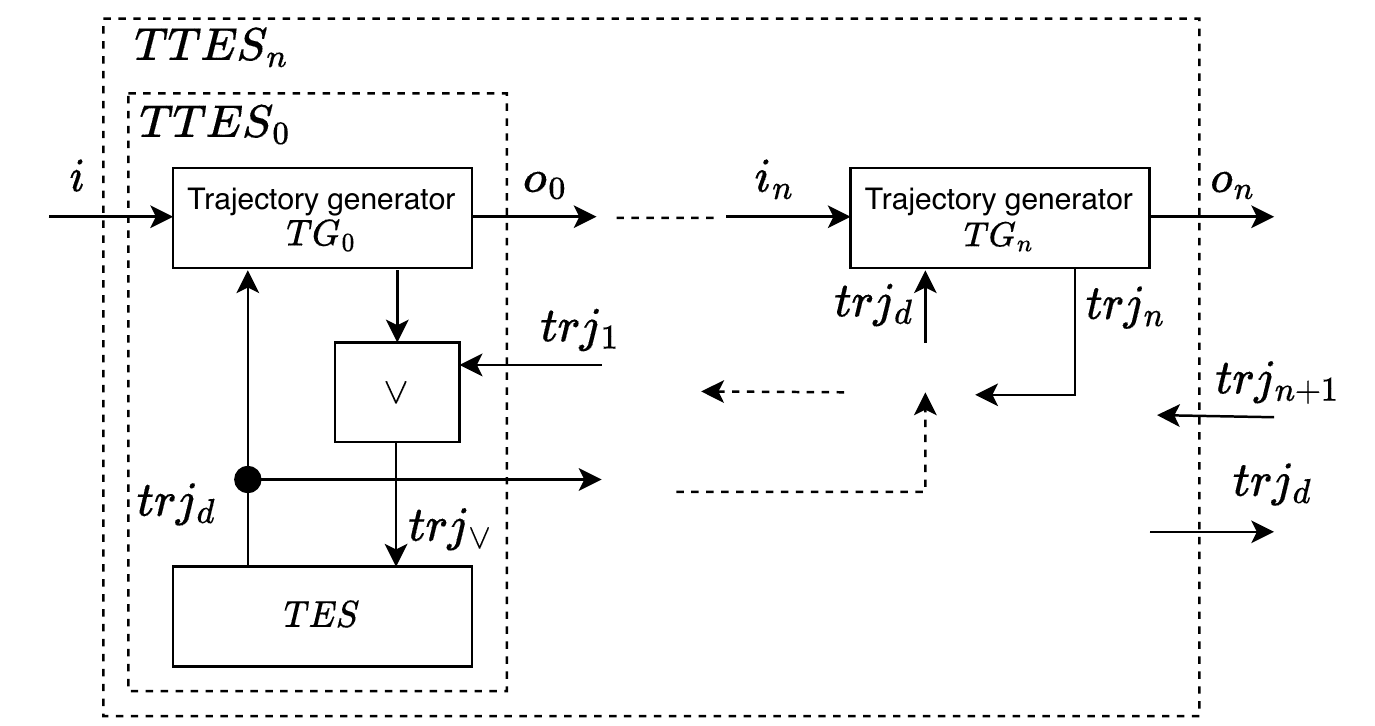}
    \caption{Recursive structure of the protocol view of $\TTES$ to prove the inductive argument.}
    \label{fig:recursive-protocol}
\end{figure}

Figure \ref{fig:recursive-protocol} shows an approach for our case study. 
The goal is to consider the generic contract $C_{\TTES_n}^{p}$ which, as shown in Figure \ref{fig:recursive-protocol} has inputs $i$ and $\vtrj_{n+1}$, and outputs $o_n$ and $\vtrj_d$, and to show that this contract is a refinement of the specification of $\TTES$. Moreover, after renaming $o_n$ to $o$ and $\vtrj_{n+1}$ to $\vtrj$, this contract should be invariant with respect to $n$. To achieve this goal, we use a generic $\TG$ component with input $i$, output $o$, and the usual interface with the $\TES$ system which comprises the two signals $\vtrj_\vee$ and $\vtrj_d$. As usual, this component assumes that $\TES$ responds to $\vtrj_\vee$ with a $\vtrj_d$. It also guarantees that an input results in one, and only one output, and that after the output is generated, all further $\vtrj_d$ signals are ignored until a new input $i$. Several instances of this $\TG$ components can be chained by connecting the output of one to the input of the next. A $\TG$ instance handles one segment of a trajectory and then passes the duty of handling the next segment to the next $\TG$ instance in the chain. For $n=0$, we have a three-component system $\{C_{\TG_0}^{(p)},C_{\TES}^{(p)}, C_{\vee}\}$ and we can show that it refines $C_{\TTES_0}^{(p)}$. Notice that $C_{\vee}$ is a contract that accepts all environments and guarantees that the output is the disjunction of its inputs. The inductive step can be shown to hold by assuming that $C_{TTES_n}^{(p)}$ holds, and showing that the system $\{C_{TTES_n}^{(p)},C_{TG_{n+1}}^{(p)},C_{\vee}\}$ refines $C_{TTES_{n+1}}^{(p)}$\footnote{Models can be found at \url{https://github.com/alessandro-pinto/icaa2023models/tree/main/TTES/protocol}}.

Refinement of the functional viewpoint follows a similar strategy. The $\TG$ component transforms a route $\vrt$ into a vector $\mathbf{\vtrj}$ of trajectory segments of length $n+1$. This vector has the favorable property that each element is a trajectory that ends where the next segment starts. Moreover, it is implicit that the required velocity is equal to zero at the start and at the end of each segment. This means that the guarantees provided by $\TES$ at the end of a trajectory (e.g., that the velocity of the aircraft is equal to zero, and that the state estimate has a bounded error), satisfy the assumptions on the initial state required by $\TES$ for the next trajectory segment. Thus, even in this case, an inductive argument can be used to show that if $\TES$ correctly executes a trajectory segment, then it correctly executes a vector of trajectory segments $\mathbf{\vtrj}$ of any length.

Modeling $\TES$ requires introducing concepts from control theory. Specification languages for this purpose include differential dynamic logic~\cite{platzerDifferentialDynamicLogic2008} which is expressive enough to capture the dynamics of the control loop. A reasoning system like KeYmaera\footnote{\url{https://keymaerax.org/}} can then be used to check whether a dynamical system satisfies certain properties. However, in the requirement generation phase, we are interested in eliciting the properties of the dynamics of a system rather than the equations that govern them. Thus, we propose an ontological approach based on a taxonomy of environments and properties that can be modeled using MS-FOL. 

Controllers typically make assumptions about the dynamics of the system they are controlling. In our taxi scenario, for example, we could make the assumption that the environment behaves as a simple car \cite{LaValle2006}. Assume that the ontology used to define environments includes a type $SimpleCar$ which represents the set of all simple car models. This model has one parameter $L : SimpleCar \rightarrow \mathbb R_{\geq 0}$, denoting the distance between the back and front wheels. Also, consider a universal function $dyn$ which maps an environment model, an initial state, and a trajectory of actuation inputs, to a trajectory of states of the environment. For reasons explained earlier in this paper, axiomatization of this function may not be needed to prove compatibility or refinement among contracts. The trajectory follower assumes that the environment behaves as a simple car, perhaps with uncertain length, for all initial conditions, and that the state estimate is close to the initial state commanded by the input trajectory. It guarantees to find a control input such that the control loop follows the given trajectory: 
\begin{equation*}
    \begin{split}
        A_{TF}^f &= 
        InBall(\hat{s}(0),trj(0)) \wedge L_l \leq L(sc) \leq L_u \wedge \\
        & \wedge \forall t \geq 0, init. 
          \ InBall(\hat{s}(t),dyn(sc,init,a)(t),\delta_{o}) \\
        G_{TF}^f &=  \forall p \in [0,|trj|]. \exists t. p= intg(t) \wedge \\
        &\wedge InBall(trj(p),dyn(sc,\hat{s}(t_0),a)(t),\delta_{c})
    \end{split}
\end{equation*}
In this model, $\mathit{sc}$ is of type $SimpleCar$ in FOL, $L_l$ and $L_u$ are uncertainty bounds on the car model, and  $\delta_o$ and $\delta_c$ are admissible estimation error and guaranteed closed-loop error, respectively. The environment makes no assumptions and guarantees to be a simple car: $G_{ENV}^f =   \forall t \geq 0, init. 
          \ InBall(\xi(t),dyn(sc,init,a)(t),\delta_{e})$.
Finally, the state estimator guarantees a bounded error: $A_{SE}^f = \forall t \geq 0. \ InBall(\xi(t),\hat{s}(s),\delta_{s})$. The specification of $\TES$ can simply be stated as follows:
\begin{equation*}
    \begin{split}
        A_{TES}^f &= InBall(\hat{s}(0),trj(0)) \\
        G_{TES}^f &= \forall p \in [0,|trj|]. \exists t. p= intg(t) \wedge \\
        &\wedge InBall(trj(p),\xi(t),\delta_{tes})
    \end{split}
\end{equation*}

An additional axiom is also needed to compose errors: $\forall p1, p2, p3, \delta_1 \geq 0, \delta_2 \geq 0. (InBall(p1,p2,\delta_1) \wedge InBall(p2,p3,\delta_2) \Rightarrow InBall(p1,p3,\delta_1 + \delta_2))$.

\paragraph{Gap analysis and research directions} Autonomous systems that are goal-driven and built to operate in a wide range of environments are free to choose the capabilities to exercise at any given time. These capabilities may have to be used an undetermined number of times. Proving that a system satisfies its requirements independently from the mission horizon may not be straightforward. We have shown an example where we have addressed this problem by proving an inductive proof of refinement. However, \emph{theoretical results for a general approach are lacking}. One possible direction towards a general approach is to consider higher-order contracts along the lines of higher-order components~\cite{Cataldo:EECS-2006-189}. These are contracts parameterized with other contracts, and that define their internal structure among their guarantees. This research could shed light on the generic properties of plans with a given structure.

We have also shown how modeling requirements for control systems could be achieved using an approach based on standard ontologies. For this approach to be viable, \emph{an extensive reusable library of requirements for efficient compositional declarative specification of control systems should be developed} (similar in nature to other libraries for mathematics developed for instance in the context of theorem provers such as mathlib4\footnote{\url{https://github.com/leanprover-community/mathlib4}}, and IsarMathLib\footnote{\url{https://isarmathlib.org}}). The selection of the level of abstraction of such ontology (i.e., the symbols and the axioms to be encoded) is a research area on its own. The level of abstraction should be high enough to limit complexity, but it should still allow the verification of relevant properties.

\subsection{Component implementation uncertainty}
\label{sec:component-uncertainty}
The central problem we focused on in this paper is refinement verification between a specification contract $C$ and the composition of a set of contracts $\{C_1,\ldots,C_n\}$. Contracts are decomposed until they are ready to be implemented. An implementation is a specification $I$ that satisfies the contract: $I \otimes A_i \leq G_i$. In some cases, an implementation can precisely satisfy a contract as in the case of protocols implemented by finite state machines in software which can be verified through various methods. 
However, this is not in general the case for autonomous systems. Some functional requirements cannot be refined into well-defined input-output relations as in the case of data-driven approaches that use machine learning techniques to approximate a function described by examples. Consider the last step of our refinement chain, where we have defined contracts for the control system. The trajectory follower imposed a maximum error in the state estimation with respect to the ideal simple car dynamics. If the GPS and IMU sensors, even with a well-designed filter, are not able to deliver a good enough level of performance, a possible approach is to rely on a visual servoing system composed of a camera and a processing algorithm. The processing algorithm  computes the error between the current aircraft pose and the ideal pose that follows the centerline, while another algorithm computes the distance between the nose of the aircraft and a hold-short line when in the field of view of the camera. 

While it is possible to specify that a component must identify the centerline and compute the cross-track error, refining such requirements into an input(image)-output(error) behavior is not. Thus, a neural network is trained with a set of examples, each comprising an image and the corresponding error (as done for example in TaxiNet \cite{8718224}). Such a network is then tested against a set of test pairs (different from the training set). The result is a score that talks about the accuracy of the implementation. In other words, among the set of components of a system, there could be one (or more) such that its implementation $I_i$ satisfies its contract only probabilistically: $P_{\geq \alpha}(I_i \otimes A_i \leq G_i)$\footnote{Due to the lack of space, we don't go into the details of this definition. The reader can imagine a probability measure over the set of environments $A_i$, and correspondingly, a resulting discrete probability distribution over the Boolean proposition $I_i \otimes A_i \leq G_i$.}. In fact, in the case of neural networks, the implementation is only partially tested, and therefore we can really only state that $P_{\geq \alpha}(I_i \otimes A_i' \leq G_i)$ for $A_i' \subseteq A_i$. It could be possible perhaps to state $P_{\geq \beta}(I_i \otimes (A_i\setminus A_i') \leq G_i)$, where $\beta \leq \alpha$, capturing the fact that there is some confidence that the neural network generalizes outside the training set (see for example \cite{katz2022verification} for an approach to component-level verification).
Thus, given that some contracts are satisfied only probabilistically, such implementation uncertainty must be propagated through the refinement chain to compute the probability that the top-level requirements are satisfied. 

\paragraph{Gap analysis and research directions.}
The specification languages used to define contracts, and the logic used to define refinement conditions as in Equation \ref{eq:contract-refinement-conditions} should provide a formal treatment of uncertainty. Probability is one way to capture uncertainty, but other frameworks should be considered such as evidence theory\cite{shaferMathematicalTheoryEvidence1976}, Bayesian networks\cite{ARACHNE}, or epistemic modal logic\cite{faginReasoningKnowledge2003}. There is a need for \emph{efficient automated reasoning tools that can solve the satisfiability problem for languages that support multiple types of uncertainty}. Finally, \emph{there is a need for results on the uncertainty associated with SA and DM capabilities of an autonomous system}. Some initial results have been recently published for geometric perception\cite{carloneEstimationContractsOutlierRobust2022a}, but more research needs to be done to provide similar results for several other domains.

\subsection{Relaxing requirements}
\label{sec:relaxing-requirements}
The current system is not capable of dealing with unexpected situations. What if we wanted to improve the system with the ability to handle unforeseen scenarios? Revision of requirements is typical in the design process which is very iterative in nature. Furthermore, the design process for autonomous systems is often agile, with requirements and implementations evolving over time as the system under design operates in its environment.  

While a compositional framework is ideal for incremental design, adding or relaxing requirements may induce radical changes to the architecture which may, in turn, require a completely new set of models. For example, consider relaxing our scenario by allowing potential unforeseen obstacles that the aircraft has to detect and avoid. This change means that the controller may not be able to follow exactly the trajectory given by the trajectory followers because the aircraft speed might have to be reduced, and the aircraft might have to stop in the middle of executing it. The trajectory follower would need to be aware of such changes. If the trajectory becomes infeasible, the route generator might have to be eventually notified. These changes in the top-level requirements ripple through the entire model which may have to undergo radical changes. For example, simple $\vtrj_d$ and $\vrt_d$ binary signals may have to be expanded to larger alphabets, or additional signals may have to be added. Additional interactions may have to be added between the state estimator and all decision-making function to maintain the state of plan execution which becomes an additional input to $\RG$ and $\TG$. Race conditions may emerge between these new signals, $\vtrj_d$, and $\vrt_d$ which would need to be handled or eliminated. These changes may lead to a complete model re-development cycle. 

\paragraph{Gap analysis and research directions.}
The merging operator\cite{Incer:EECS-2022-99} on contracts enables incremental requirement development and design. However, adding requirements may lead to radical changes in the architecture of a system and, consequently, radical changes in the models used for verification. Given that the modeling effort is generally high and model reuse is key to efficiency, this situation should be avoided. \emph{Further research is needed for the development of general model re-usability guidelines and standard architectures for autonomous systems that are robust against changes in requirements}.

\section{Conclusions}
\label{sec:conclusions}
We have used a contract-based framework to model an autonomous aircraft taxi system. We have leveraged the ability to model a system along different views in order to use the most appropriate language and efficient automated reasoning tool for each. We have shown how a model can be structured in a top-down fashion from requirements to low-level control through a series of refinement steps. We have also suggested a way to capture requirements for the lower-level control functions via an abstraction into an ontology of known properties that can then be verified by other dedicated tools. We have identified several research directions including the development of a methods to orchestrate different analysis methods, a comprehensive set of libraries of efficient formal models for autonomous systems, methods to explain results generated by formal method tools to subject matter experts, methods to reason about uncertain contracts for imperfect implementations, and standard architectures for model reuse.

\bibliographystyle{IEEEtran} 
\bibliography{refs}

\end{document}